\documentstyle[10pt,psfig]{espart}
\begin{document}
\begin{frontmatter}
  \title{{\small Physica A, Vol.216, 199-212 (1995)}\\[0.5cm]
Simulation of rotating drum experiments using non--circular
    particles} \author[HUB]{Volkhard Buchholtz\thanksref{bylinevb}},
  \author[AND]{Thorsten P\"oschel\thanksref{bylinetp}}\\ and
  \author[KFA]{Hans-J\"urgen Tillemans\thanksref{bylinehjt}}

  \address[HUB]{Humboldt--Universit\"at zu Berlin, Institut f\"ur
    Physik, Unter den Linden 6,\\ D--10099 Berlin, Germany}

  \address[AND]{Arbeitsgruppe Nichtlineare Dynamik, Universit\"at
    Potsdam, Am Neuen\\ Palais, D--14415 Potsdam, Germany.
    http://summa.physik.hu-berlin.de:80/$\sim$thorsten/}

  \address[KFA]{H\"ochstsleistungsrechenzentrum, Forschungszentrum
    J\"ulich, D--52425 J\"ulich, Germany and\\
    Universit\"at zu K\"oln, Institut f\"ur Theoretische Physik,
    Z\"ulpicher Str. 77, D--50997 K\"oln} \date{\today}
  \thanks[bylinevb]{E--mail: volkhard@itp02.physik.hu--berlin.de}
  \thanks[bylinetp]{E--mail: thorsten@hlrsun.hlrz.kfa--juelich.de}
  \thanks[bylinehjt]{E--mail: hjt@hlrsun.hlrz.kfa--juelich.de}
\begin{abstract}
  We investigate the flow of granular material in a rotating cylinder
  numerically using molecular dynamics in two dimensions. The
  particles are described by a new model which allows to simulate
  geometrically complicated shaped grains. The results of the
  simulation agree significantly better with experiments than the
  results which are based on circular particles.
\end{abstract} 
\end{frontmatter}

\section{Introduction}
The flow of granular material in a rotating cylinder or drum reveals
many interesting effects and has been studied by physicists
experimentally as well as theoretically in various articles. Moreover,
the knowledge of the dynamics of the material is of high interest in
engineering since many important technological processes are based on the
movement of granular material in rotating cylinders, such as mixing
and milling (see
e.g.~\cite{Rolf:1993KUG,Rothkegel:1992,HashimotoWatanabe}).

Nakagawa used magnetic resonance imaging (MRI, NMR) to observe mustard
seeds moving in a rotating drum~\cite{Nakagawa:1993ug}. Seeds are
suited because their oil contains free protons which are necessary for
this method. The velocity and the concentration of the seeds in the
drum could be determined very elegantly with this method. Woodle and
Munro~\cite{WoodleMunro:1993} and Raj\-chen\-bach~\cite{Rajchenbach:1990}
found experimentally a hysteresis for the transition from stick--slip
motion of the particles to continuous motion when changing the
rotation velocity of a rotating drum. There are two different critical
angular velocities which separate the regimes of stick--slip and
continuous flow. If the rotation frequency is increased the critical
angular velocity is higher than in the case of decreasing frequency.
Raj\-chen\-bach found experimentally that the relation between the surface
current $J$, which is proportional to the angular velocity of the
rotating drum, and the surface angle of the granular medium $\Theta$
obeys a power law~\cite{Rajchenbach:1990}:
\begin{equation}
J \sim (\Theta - \Theta_c) ^ m .
\end{equation}
For the exponent $m$ they measured a value of about 0.5. Rolf and
Rothkegel used a very interesting and sophisticated experimental setup
to examine the collision frequency between particles in a rotating
drum~\cite{Rolf:1993KUG,Rothkegel:1992}. Some of the particles
contained equipment for pressure measuring and microelectronic devices
to store the data or transmitters to transmit the data to an external
computer. Several regions with different collision rates were found.

Presently there are two different methods available for the numerical
simulations of these effects.
\begin{enumerate}
\item A method which is based on a model by Visscher and
Bolsterli~\cite{VisscherBolsterli:1972ug} was introduced in 1992 by
Jullien et al. to examine a shaken granular
medium\cite{JullienMeakinPawlovitch:1992}. Here the dynamics of the
particles is calculated separately for each particle. In the case of
the shaken box the particles are ordered according to their distance
from the bottom of the box. Then one calculates the motion of the
particles beginning with the one which has the lowest distance to the
bottom while all other particles are fixed. The particle moves until
it reaches a local minimum where it gets stuck. Then one calculates
the next particle while all others are at fixed positions and so on. This
method is very fast but it has the disadvantage that inertia and
elasticity are neglected and only particles with a restitution
coefficient of zero can be simulated. For this reason the model has
been attacked in ref.~\cite{MehtaBarkerGrimson:1993ug}.  Nevertheless
under certain conditions this model is able to reproduce the
experimental behavior correctly. With a model which is based on the
same method Baumann et al. found an interesting phenomenon which is
also measured
experimentally~\cite{BaumannJobsWolf:1993,BaumannJanosiWolf:1994,BaumannJanosiWolf:1995}:
Two sorts of circular particles, originally perfectly mixed, will
decompose after only one or two rotations of the drum. The smaller
disks gather themselves in the center of the drum.

\item There are several models for the simulation of the dynamics of
granular material using the molecular dynamics concept
(e.g.~\cite{AllenTildesley:1987}). Most of the authors apply a model
introduced by Cundall and Strack~\cite{CundallStrack:1979} and Haff and
Werner~\cite{HaffWerner:1986} where it is assumed
that the particles are ideal circular disks $i$ with radii $R_i$ which
interact when they touch each other. If the distance between their
centers $\vec{r}_i$ is smaller than the sum of their radii they feel a
repulsive force. To be able to capture the loss of energy and to
simulate a restitution coefficient this movement is damped.  Hence the
particles feel a force $\vec{F}_{ij}$ with the normal component
$F_{ij}^N$ and the shear component $F_{ij}^S$
\begin{equation}
  \vec{F}_{ij}= F_{ij}^N \cdot \frac{\vec{r}_i-\vec{r}_j}{\left|
    \vec{r}_i-\vec{r}_j \right| } + F_{ij}^S \cdot \left(
  \begin{array}{cr} 0 & -1 \\ 1 & 0 \end{array} \right)
    \frac{\vec{r}_i-\vec{r}_j}{\left| \vec{r}_i-\vec{r}_j \right| }
\end{equation}
with 
\begin{eqnarray}
  F_{ij}^N &=& Y \cdot \left( \left| \vec{r}_i-\vec{r}_j\right| - R_i
  - R_j \right)^{\frac{3}{2}} - m_{ij}^{eff} \cdot \gamma_N \cdot
  v_{rel}^{N}
\label{fnormal}\\ 
F_{ij}^S &=& - min( m_{ij}^{eff} \gamma_{S} |v_{rel}^{S}|, \mu
|F_{ij}^{N}|)
\label{ftang} \\
\vec{v^{rel}}&=& \vec{v_j} - \vec{v_i} \label{surfvelocity}\\ 
m_{ij}^{eff} &=& \frac{m_i \cdot m_j}{m_i + m_j} \label{meff}
\end{eqnarray}
with the Young modulus $Y$, the damping coefficients $\gamma_N$ and
$\gamma_S$ and the Coulomb friction coefficient $\mu$.
Eq.~(\ref{surfvelocity}) describes the relative velocity of the
surfaces of the particles at the point of contact and eq.~(\ref{meff})
gives the effective mass. Eq.~(\ref{ftang}) takes the Coulomb friction
law into account, saying that two particles slide on top of each other
if the shear force overcomes $\mu$ times the normal force.
Eq.~(\ref{fnormal}) comes from the Hertz law \cite{Hertz:1882ug} for
the force between two rigid bodies which are in contact with each
other.  This model has been applied to simulations of rotating drums
by several authors. Ristow investigated the particle size segregation
in a rotating drum in two dimensions
\cite{Ristow:1994,RistowCantelaubeBideau:1994ug}, P\"oschel and
Buchholtz investigated the irregularities of the particle
stream~\cite{PoeschelBuchholtz:1993CSF}.
\end{enumerate}

There are some models to simulate more complex shaped grains
which are composed of circular disks. In the model by Gallas and
Soko{\l}owski \cite{GallasSokolowski:1993} the grains consist of two
disks connected with each other by a stiff bar.
Walton and Braun used a three dimensional MD simulation with spheres
and with more complicated two--dimensional particles consisting of four or eight
circular disks rigidly connected with each other~\cite{WaltonBraun:1993}.
With this model they examined the transition from stationary to
sliding and the transition from sliding to raining flow in a rotating
drum. P\"oschel and
Buchholtz~\cite{PoeschelBuchholtz:1993,BuchholtzPoeschel:1994PA}
describe grains built up of five circular disks where one of them is located
in the center of the grain and four identical disks are at the
corners of a square. Each pair of neighboring disks is connected by
a damped spring. The latter model was applied to the rotating cylinder
\cite{PoeschelBuchholtz:1993CSF} and it was shown that the simulation
results agree much better with the experiment than equivalent
simulations using circular disks. Especially it was shown that one can
reproduce stick--slip motion and avalanches which was not possible
with circles. The inclination and the dependency of the inclination
on the angular velocity, however, did not agree well with the
experiment.

Another model for non--circular grains has been proposed by Mustoe
and DePorter~\cite{MustoeDePorter:1993}. There the boundary geometry
of the particles is defined (in local coordinates) by
\begin{equation}
f_i(x_i,y_i)=\left[\frac{\left| x_i \right|}{a_i}\right]^{n_i} + 
             \left[\frac{\left| y_i \right|}{b_i}\right]^{n_i} - 1 = 0.
  \label{hogue.eq}
\end{equation}
By changing $n_i$ from 2 to $\infty$ the shape of the particles varies
continuously from elliptical to rectangular.
They use a complicated algorithm to detect collisions between
particles.  Hogue and Newland~\cite{HogueNewland:1993} investigate the
flow of granular material on an inclined chute and through a hopper
using a model where the boundary of the convex particles is given by a
polygon with up to 24 vertices.  To detect whether two particles touch
each other one has to calculate the intersections between each pair of
vertices. During collisions energy is dissipated according to
Stronge's energy dissipation hypothesis~\cite{Stronge} in normal
direction, whilst Coulomb's friction law models the energy losses in
tangential direction.

Using grains which consist of interconnected circles or of particles
described by eq.~(\ref{hogue.eq}) it is not possible to simulate
particles with sharply formed corners. For some effects it seems to be
essential to simulate such particles to reproduce the experimental
observed effects. This point is discussed in detail
in~\cite{BuchholtzPoeschel:1994PA,TillemansHerrmann:1995,Tillemans:1994}.
In this paper we present molecular dynamics simulations where the
particles are simulated using a more sophisticated model.

\section{The model}
\begin{figure}[ht]
  \centerline{\psfig{figure=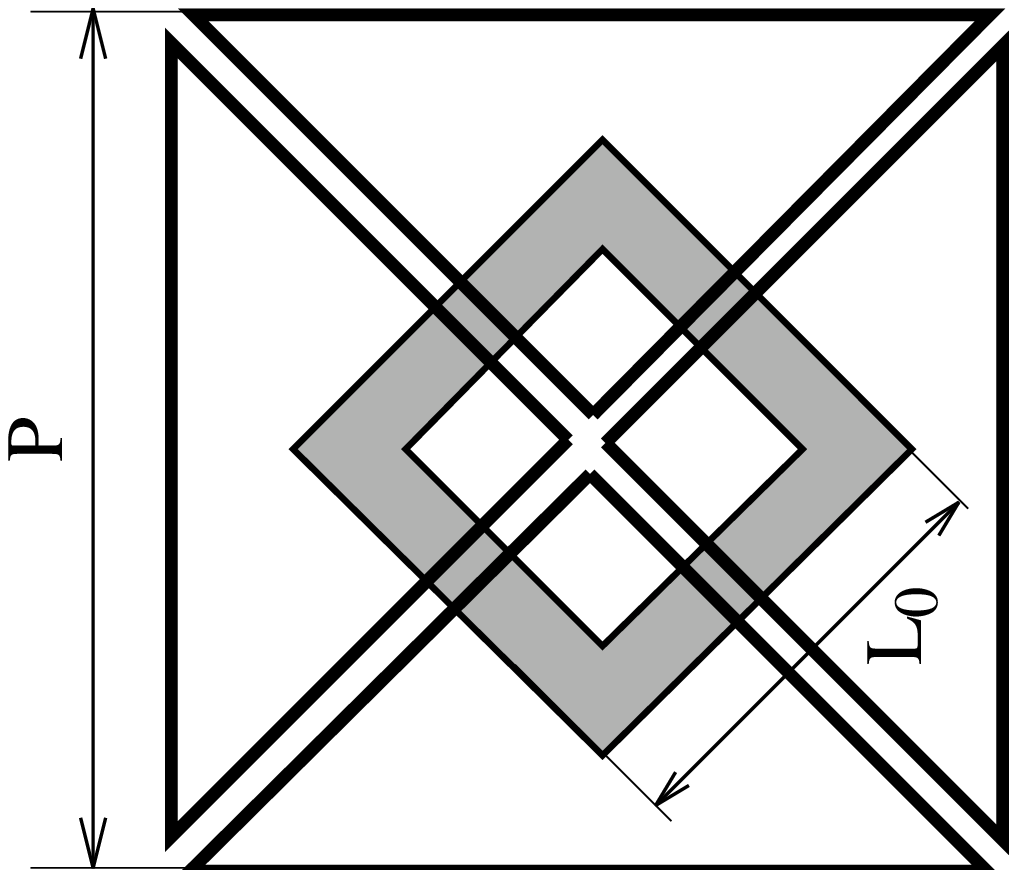,width=4cm,angle=270}
    \hspace{0.5cm}
    \psfig{figure=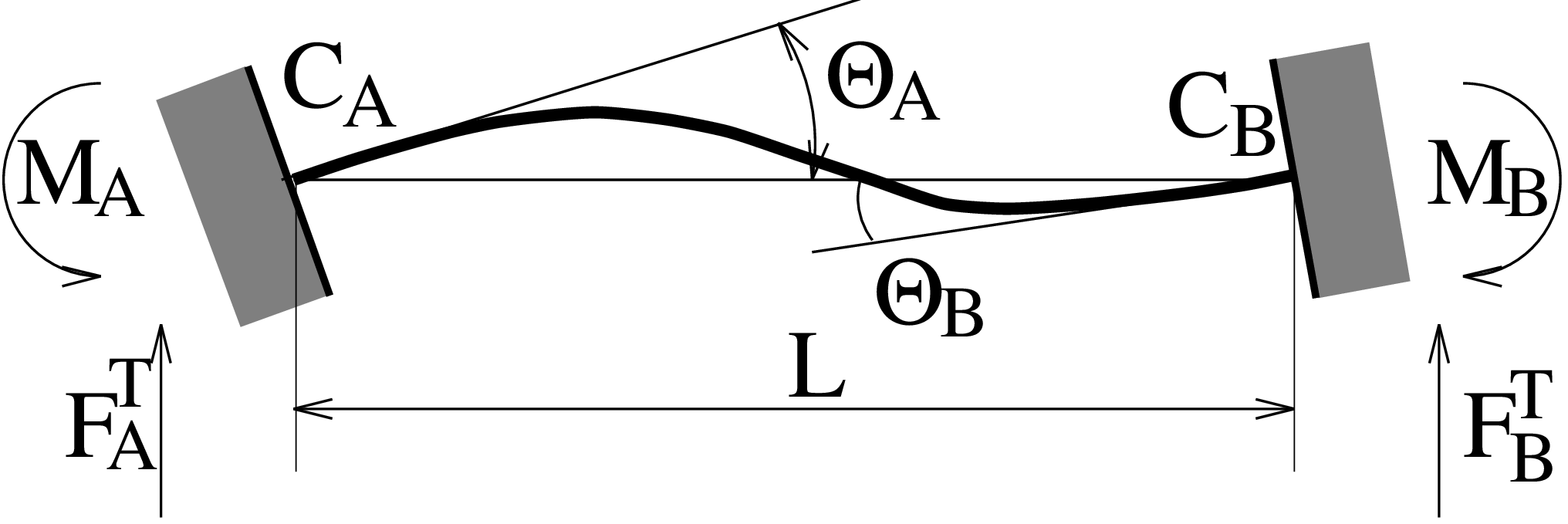,width=8cm,angle=270}}
  \vspace{0.5cm}
\caption{\em Draft of a complex particle (left). The four triangles are
  connected by damped leaf springs as shown in the right figure. The
  beams are fixed at the centre of mass points of the triangles $C_A$
  and $C_B$. When the triangles undergo collisions the beams are
  elongated and bent. They disspipate energy proportionally to the
  deformation rate.}
\label{particle.fig}
\end{figure}

The particles in our simulation consist of four triangles which are
connected by beams of length $L_0=P\cdot\frac{\sqrt{2}}{3}$ to simulate a square particle of
side length $P$ as shown in fig.~\ref{particle.fig}~(left). The ends
of the beam are fixed at the center of mass point of the connected
triangles in the direction in perpendicular to the neighboring sides of
the triangles when the grain is in its rest state, i.e. no forces are
applied. The beams are shadowed in fig.~(\ref{particle.fig}). A beam in our sense is an elastic bar which is subjected to
forces in the direction of its axis and transverse to its axis, i.e.
to normal and shear forces, and to torques acting on its ends. During
its deformation a beam dissipates energy similar to a linearly damped
spring, i.e.  proportional to its deformation rate.

The deformation of a beam between the triangles $A$ and $B$ with
centre of mass points $C_A$ and $C_B$ is completely described by the
angles $\Theta_A$ and $\Theta_B$ and by its length $L$ (see fig.~\ref{particle.fig}, right).
A deformed beam acts on the triangles with the torques
\begin{eqnarray}
  M_A &=& \frac{4EI}{L}~\Theta_A + \frac{2EI}{L}~\Theta_B ,
  \label{beammomente1}
\\ M_B &=&
\frac{2EI}{L}~\Theta_A + \frac{4EI}{L}~\Theta_B . \label{beammomente2}
\end{eqnarray}
and with the forces in perpendicular to $\overline{C_AC_B}$
\begin{eqnarray}
  F_A^T &=& \frac{6EI}{L^2}~\left( \Theta_A + \Theta_B\right),  \\ F_B^T
  &=& -\frac{6EI}{L^2}~\left( \Theta_A + \Theta_B\right) .
\end{eqnarray}
where $I$ is the moment of inertia of the beam and $E$ is the
elastic constant of the beam material. In the case of a beam with
rectangular cross section of side lengths $S$ and unity one finds
\begin{equation}
I= \frac{1}{12} S^3 . 
\end{equation}
In the direction of $\overline{C_AC_B}$ the beam acts with the forces
\begin{eqnarray}
  F_A^N &=& E\cdot \left(L-L_0 \right)\label{beamnormal1} ,\\ F_B^N &=&
  -E\cdot \left(L-L_0 \label{beamnormal2} \right) , \\ L_0 &=&
  \frac{\sqrt{2}}{3}~P .
\end{eqnarray}
The torques and forces given by
eqs.~\ref{beammomente1}--\ref{beamnormal2} are valid for small
deformations of the beam, i.e. we assume a linear superposition of
all forces and torques \cite{TimoshenkoGere:1972}. For a detailed
derivation of the formulae see \cite{PoeschelBuchholtz:1995}.

The deformation of the beams is damped proportionally to the
deformation rates
\begin{eqnarray}
  M_A^{(d)} &=& -\frac{\gamma~I}{L} ~\dot{\Theta}_A , \\ 
  M_B^{(d)} &=& -\frac{\gamma~I}{L} ~\dot{\Theta}_B , \\ 
  F_A^{N(d)} &=& -\gamma\left[ \vec{v}_A \cdot
    \stackrel{\longrightarrow}{C_AC_B} - \vec{v}_B  
    \cdot \stackrel{\longrightarrow}{C_AC_B} \right] , \\
  F_B^{N(d)} &=& \gamma\left[ \vec{v}_A \cdot 
    \stackrel{\longrightarrow}{C_AC_B} - \vec{v}_B 
    \cdot \stackrel{\longrightarrow}{C_AC_B} \right]~,
  \label{dampingmom}
\end{eqnarray}
where $\gamma$ is the damping coefficient of the beam material.  As
mentioned above the deformation of the beam is in linear approximation
completely determined by the angles $\Theta_A$ and $\Theta_B$ and the
length $L$. Hence there is no damping force acting in shear direction.

\begin{figure}[ht]
  \centerline{\psfig{figure=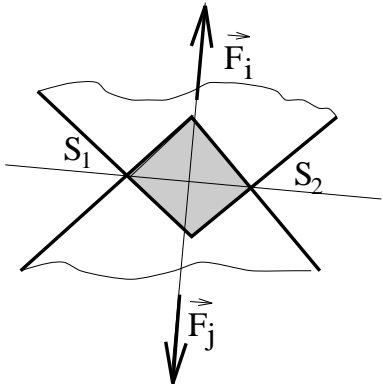,width=5cm,angle=270}}
  \vspace{0.5cm}
\caption{\em According to the Poisson hypothesis the force between two
  interacting particles is directed in perpendicular to the
  intersection line $\overline{S_1S_2}$.}
\label{force.fig}
\end{figure}

When two triangles of {\em different} grains touch each other
(Fig.~\ref{force.fig}), i.e. if there is an overlap between both they
feel a force in perpendicular to the line between the intersection
points $\overline{S_1S_2}$ according to the Poisson hypothesis
\cite{Poisson}. The absolute value of the force is assumed to be
proportional to the intersection area $Q$ times Young modulus $Y$
\begin{equation}
  \left| F_A^{is} \right| = Y~Q~.
  \label{Fdreieck}
\end{equation}
We want to point out that during the collision of the triangles the
energy is conserved. The method is described in more detail in
\cite{PoeschelBuchholtz:1995}.

Adding up all forces and torques described above one finds the net
forces and torques acting upon each particle. The dynamics of the
grains will then be calculated by a molecular dynamics scheme (see
e.g. \cite{AllenTildesley:1987}). We have chosen a Gear predictor
corrector method of 5th order \cite{Gear:1966ug}.  A beam model very
similar to ours is described in detail in
\cite{HerrmannHansenRoux:1989}.

\section{The experimental setup}
\begin{figure}[ht]
\centerline{\psfig{figure=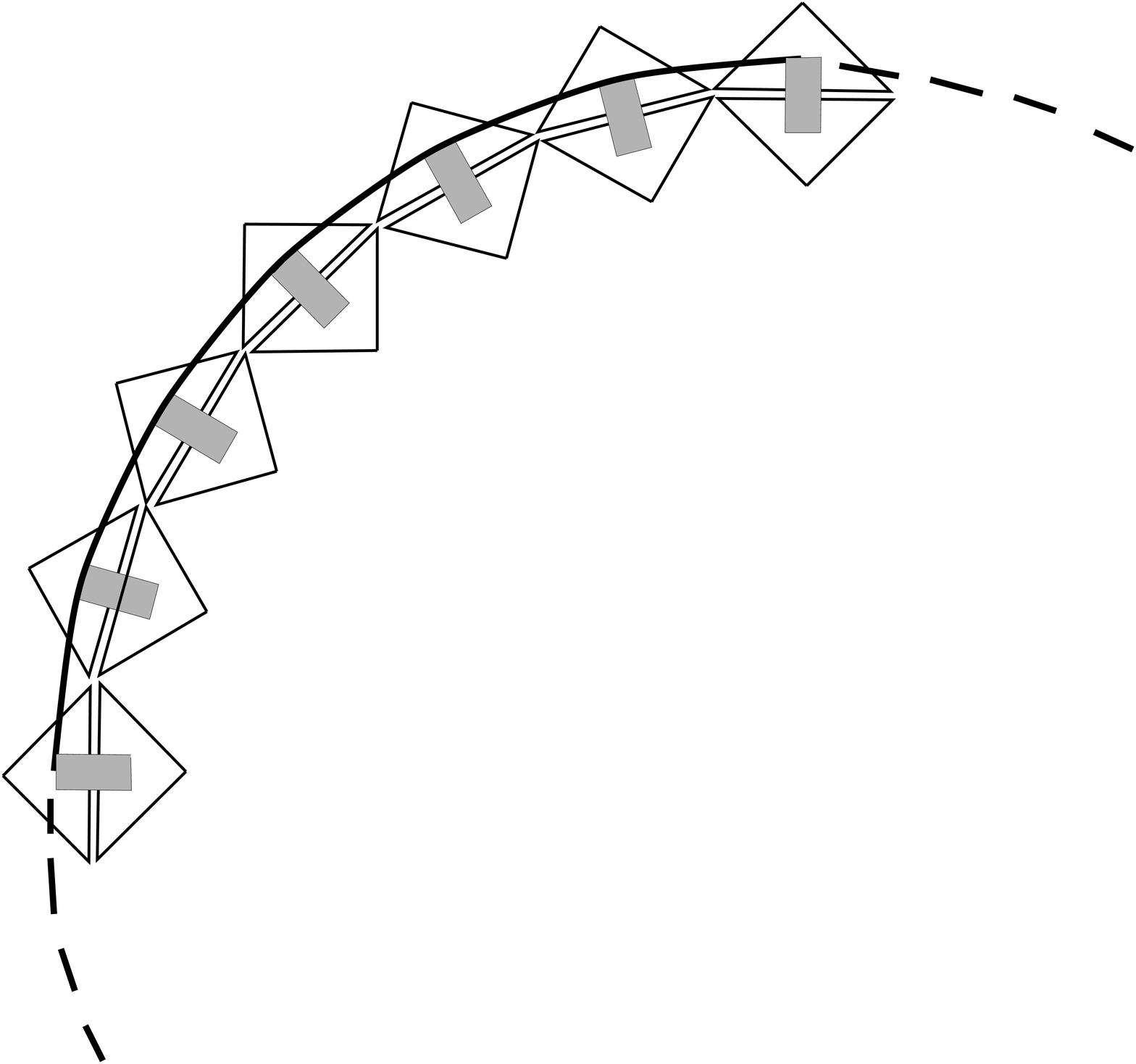,width=5cm,angle=270}}
\vspace{0.5cm}
\caption{\em The wall of the cylinder was built up of particles
  consisting of two triangles connected by damped beams to simulate a
  rough damping wall.}
\label{wall.fig}
\end{figure}

Throughout our simulations we used square particles of side length $P$
consisting of four triangles as shown in fig.~\ref{particle.fig} where
$P$ is equally distributed in the interval $P\in\left[0.1\,cm;0.2\,cm
\right]$. The Young modulus $Y$ of the material is $Y=2\cdot 10^7 g
\cdot cm^{-1} \cdot sec^{-2}$. Each connection between the triangles
consists of beams with elastic constant of the beam material $E=10^5 g
\cdot sec^{-2}$ and damping constant $\gamma=5 g \cdot sec^{-1}$.  The
moment of inertia of the beam was $I=10^{-4}\,cm^3$. When simulating
relatively stiff beams we need a very small time step $\delta t$ for
the integration of the equations of motion to guarantee numerical
stability, and hence the simulation becomes inefficient. An
alternative method is to simulate each beam by two identical beams of
lower elastic constant $E$ and using a larger time step for the
integration. We found that the latter approach works more efficiently.
The parameters have been chosen to give the best agreement of the
numerical simulated grain movement with the typical behavior of
granular media. We generated an animated sequence of snapshots of the
simulation and compared the behavior of the granular material with
what we would expect from a real material. The cylinder had a diameter
$D=4\,cm$. The wall of the cylinder consists of particles as shown in
Fig.~\ref{wall.fig} to simulate a rough damping surface. The outer
triangles are fixed at the uniformly rotating ring, the inner ones
obey the laws derived above (eqs.~\ref{beammomente1}--\ref{Fdreieck})
when they undergo collisions with freely moving particles.

With the integration step $\delta t=2\cdot 10^{-5}~sec$ the simulation
behaves numerical stable, i.e. the results do not vary when enlarging
the step a few per cent.

\begin{figure}[ht]
\centerline{\psfig{figure=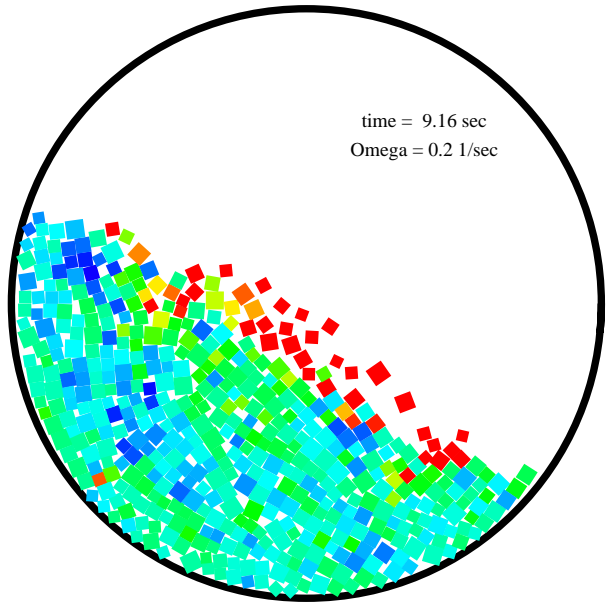,width=8cm,angle=0}}
\caption{\em Snapshot of the simulation with $N=500$ complex
  particles with $P\in [0.1\,cm; 0.2\,cm]$ in a rotating cylinder with
  diameter $D=4\,cm$. The angular velocity is $\Omega=0.2\,sec^{-1}$.
  The wall particles have not been drawn. The snapshot has been taken
during an avalanche. The gray scale codes for the particle velocity.}
\label{snap.fig}
\end{figure}

Fig.~\ref{snap.fig} shows a snapshot of the simulation with the
angular velocity $\Omega=0.2\,sec^{-1}$. The $N=500$ complex particles
consist each of $4$ triangles connected by $8$ beams. The gray scale
codes for the particle velocity, black means high velocity.
The computations have been performed on a IBM 370/R6000 workstation.
The total CPU time for the results presented in this paper was
approximately $2100$ hours.

\section{Results}
\begin{figure}[ht]
%\psdraft
\centerline{\psfig{figure=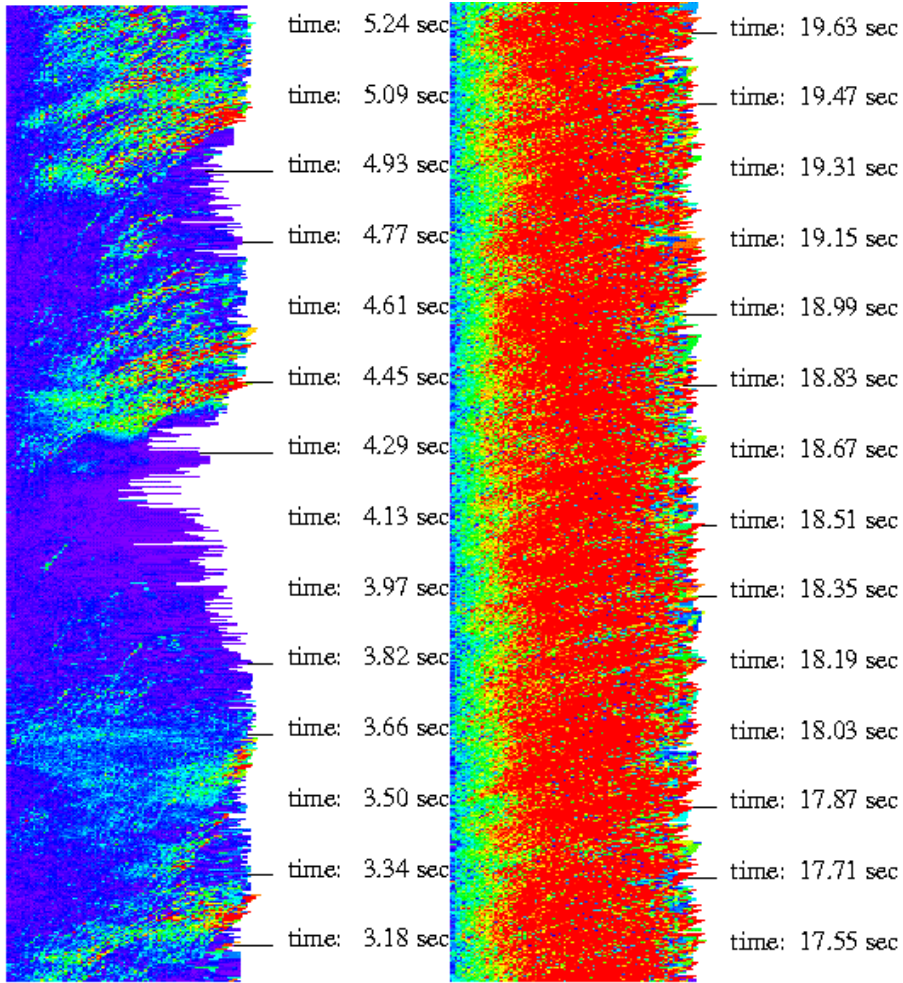,width=12cm,angle=0}}
%\psfull
\vspace{0.5cm}
\caption{\em The view at the surface of the moving granular material
  for consecutive times for $\Omega=0.1\,sec^{-1}$ (left) and
  $\Omega=1.3\,sec^{-1}$ (right). The colors encode for the particle
  velocity at the surface, red color means high velocity, blue means
  low velocity. Time rises from bottom to top. For low angular
  velocities one finds sharply distinguishable regions in space and
  time of different velocities which corresponds to the stick--slip
  like motion of the particles. The flow for $\Omega=1.3\,sec^{-1}$ is
  more homogeneous. The material flows from left to right in each
  column.}
\label{flow.fig}
\end{figure}

In our simulation we started with the angular velocity
$\Omega=0.1\,sec^{-1}$.  As we will show below the flow of the
material behaves stick--slip like for this velocity. After a
relaxation time $t_{r} = 1.5\,sec$ we began to measure the interesting
values. These values have been averaged over half a rotation before
increasing the angular velocity by $\Delta \Omega = 0.1\,sec^{-1}$.
After reaching the velocity $\Omega = 1.3\,sec^{-1}$ where the flow
moves continuously we decreased the angular velocity in the same
manner while measuring and averaging the interesting data.
Fig.~\ref{flow.fig} shows the color encoded velocity distribution of the
particles moving downwards on the surface of the material. The left
figure shows the distribution for $\Omega=0.1\,sec^{-1}$, the right for
$\Omega=1.3\,sec^{-1}$. Each horizontal line displays the values for a
certain time. Time is increased from bottom to top. Red color encodes
for high particle velocity, blue encodes for low velocity.  The color
coding is the same in the left and the right part of the figure. In
the left part one can see that for low angular velocity the particles
at the surface of the material move in a very irregular manner. One observes
long periods encoded by blue color where almost no material moves
downwards, sharply separated from short red regions where the
material moves rapidly due to avalanches. We want to call this behavior
stick--slip motion. In the equivalent figure for higher angular
velocities one does not observe avalanches. The behavior is not
stick--slip like but continuous in time and space.

In the natural as well as in the numerical experiment one observes
avalanches. The size and the time intervals between avalanches vary
irregularly. Unfortunately our numerical data are far from sufficient
to calculate reliable size distributions of the avalanches.

\begin{figure}[ht]
\centerline{\psfig{figure=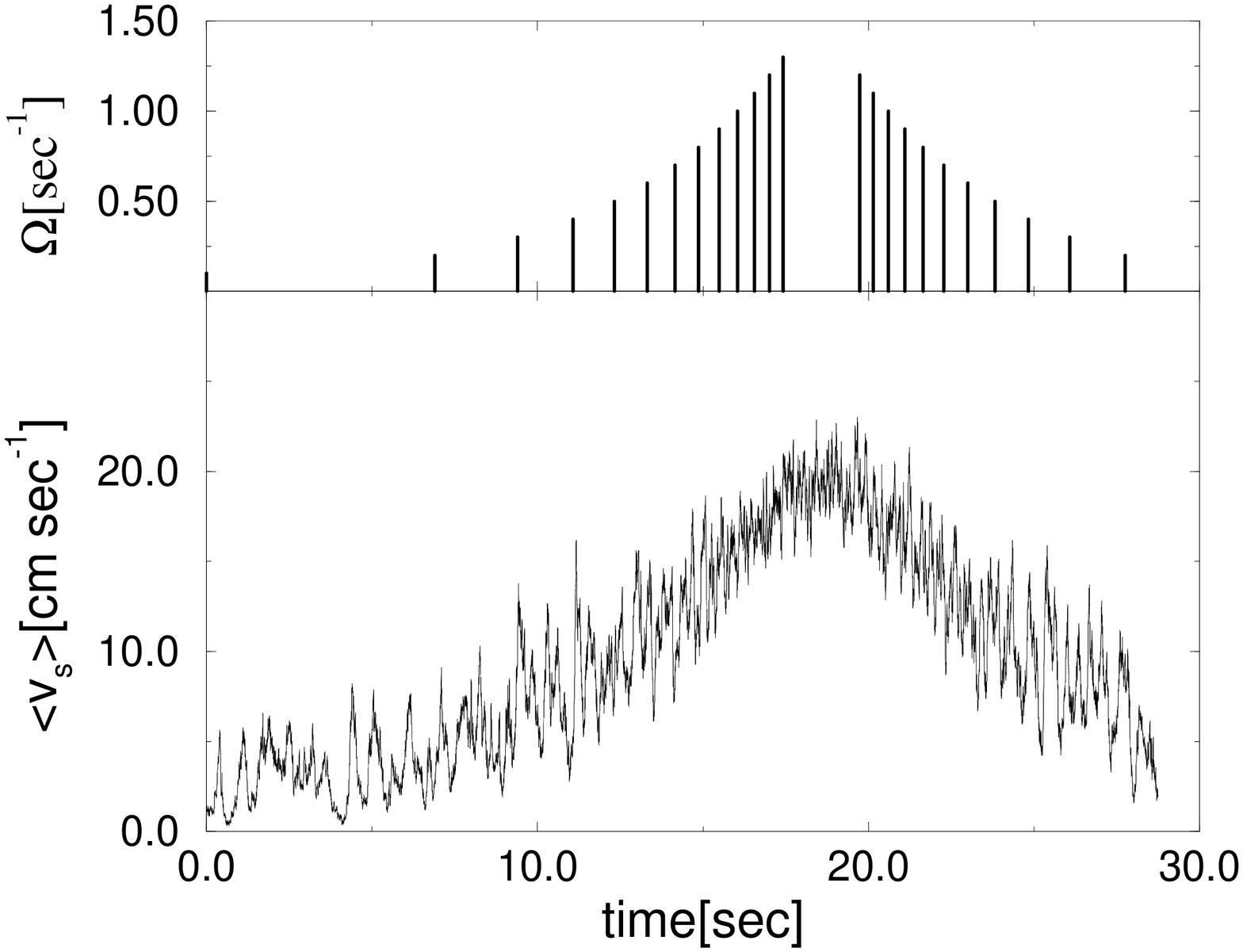,width=10cm,angle=270}}
\vspace{0.5cm}
\caption{\em The averaged positive particle velocity $\langle v_s
  \rangle$ over time (bottom figure). Each half rotation the angular
  velocity of the cylinder (top figure) was changed. For high $\Omega$
  the relative fluctuations become small compared with the relative
  fluctuations for smaller $\Omega$. }
\label{vovertime.fig}
\end{figure}

In fig.~\ref{vovertime.fig} we have drawn the averaged velocity
$\langle v_s \rangle$ of the particles moving at the surface of the
granular material over time (bottom figure). Each half rotation the
angular velocity of the cylinder (top figure) was changed according to
$\Omega = \Omega+ \Delta \Omega$ while accelerating, and $\Omega =
\Omega- \Delta \Omega$ while decelerating. The fluctuations become
smaller since the flow is much less irregularly for larger rotation
velocity. For the case of low rotation velocity one finds avalanches
due to the stick--slip characteristics while for high rotation speed
the flow becomes more homogeneous.

\begin{figure}[ht]
\centerline{\psfig{figure=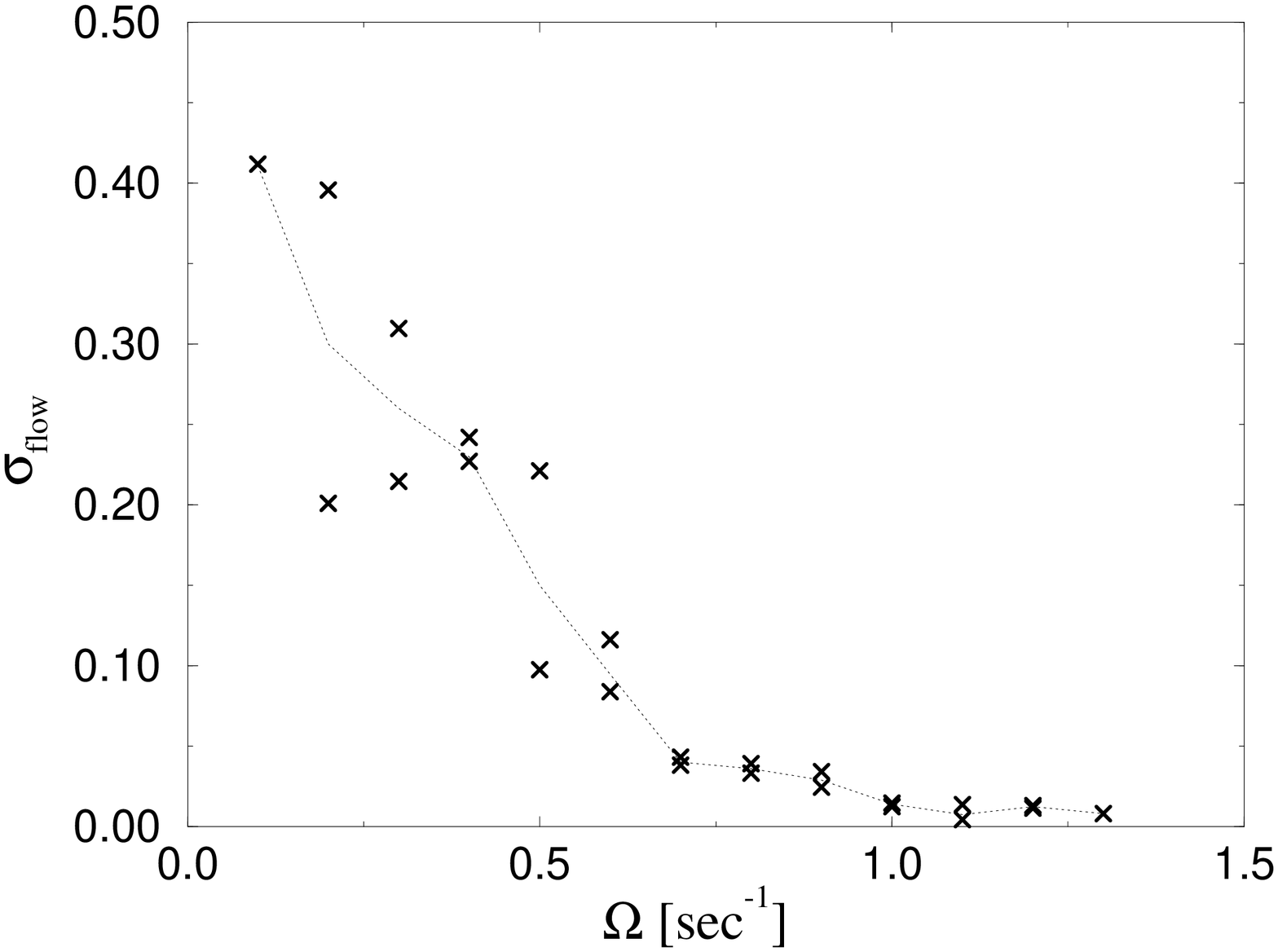,width=10cm,angle=270}}
\vspace{0.5cm}
\caption{\em The relative standard deviation 
(eq.~\ref{stdev.eq}) of the time series of the positive material flow
  in the rotating cylinder. For each $\Omega$ we measure two values
  $\sigma_{flow}$ according to our acceleration--deceleration
  schedule. The dashed line connects the mean values of each of these
  pairs. We observe a transition between the two different flow
  regimes, but we are not able to determine a precise transition point.
  The transition takes place at the rotation velocity $\Omega \approx
  0.6\,sec^{-1}$.}
\label{stdev.fig}
\end{figure}

In the simulation we recorded the time series of the material flow
gliding down at the surface of the granular material $f_\rightarrow$.
When we draw the relative standard deviation
\begin{equation}
\sigma_{flow}= \sqrt{\frac{\overline{(
        \overline{ f_\rightarrow } - 
         f_\rightarrow )^2}}
        {\overline{ f_\rightarrow }^2}}
  \label{stdev.eq}
\end{equation}
of this time series as a function of the rotation velocity $\Omega$
(fig.~\ref{stdev.fig}) we find that for $\Omega > \Omega^{cr} \approx
0.6\,sec^{-1}$ the relative standard deviation $\sigma_{flow}$ is less
than 0.1. For smaller $\Omega$ the standard deviation
$\sigma_{flow}$ rises dramatically due to the transition from the
continuous regime to the stick--slip motion. For each value $\Omega$
we find two values $\sigma_{flow}$, one when accelerating the cylinder
and one when decelerating. The dashed line in fig.~\ref{stdev.fig}
connects the mean values of each of these pairs.

Our acceleration--deceleration schedule was intended to reproduce the
hysteresis in the critical transition point $\Omega^{cr}$ where the
characteristics of the flow changes suddenly, which can be observed in
the experiment~\cite{Rajchenbach:1990}. When accelerating the cylinder
the transition occurs at the velocity $\Omega^{cr}_\uparrow$ which is
higher than the transition point $\Omega^{cr}_\downarrow$ when
decelerating the cylinder. From the results presented so far we cannot
observe a sharp transition between the stick--slip flow and the
continuous regime as it is observed in the
experiment~\cite{Rajchenbach:1990}. Hence we are not able to reproduce
the hysteresis described above.

\begin{figure}[ht]
\centerline{\psfig{figure=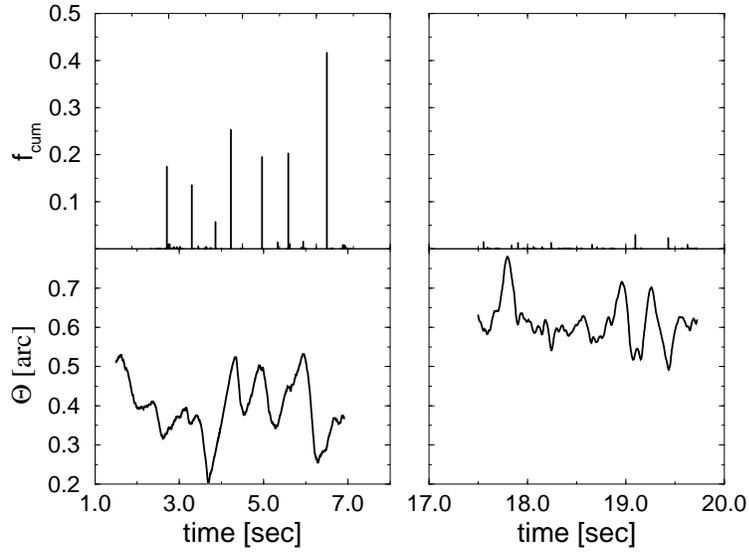,width=10cm,angle=270}}
\vspace{0.5cm}
\caption{\em Time series of the inclination $\Theta$ of the material 
  over the time (lower figures) for the angular velocities
  $\Omega=0.1\,sec^{-1}$ (left) and $\Omega=1.3\,sec^{-1}$ (right).
  The values at the time axis correspond to the previous figures. In
  the first case the angle varies more irregularly due to avalanches.
  The upper figures show the material flow (explanation in the text).}
\label{phiovert.fig}
\end{figure}

Our system is too small to provide a smooth surface to measure the
inclination $\Theta$ directly. Therefore we applied an indirect method
which is explained in detail in an earlier
paper~\cite{PoeschelBuchholtz:1993CSF}.  In fig.~\ref{phiovert.fig} we
have drawn the inclination $\Theta$ of the surface of the material for
low angular velocity $\Omega=0.1\,sec^{-1}$ (left) and for higher
velocity $\Omega=1.3\,sec^{-1}$ (right).  The values at the time axis
correspond to those in figs.~\ref{flow.fig}~and~\ref{vovertime.fig}.
In the first case the angle is smaller and varies much more
irregularly due to avalanches. The mass of the material remains
constant throughout the simulation. Hence the averaged flow over time
in positive horizontal direction $\overline{f_\rightarrow}$ which is
due to material flow at the surface equals the flow in negative
direction $\overline{f_\leftarrow}$ given by the continuous material
transport due to the rotation of the cylinder. The net flow $\Delta f=
f_\rightarrow - f_\leftarrow$ fluctuates around zero. When the
material moves stick--slip like in a clockwise rotating cylinder we
expect that there are relatively rare events (avalanches) when the
material moves in positive direction and time intervals, where the
flow on the surface is almost zero, i.e. $\Delta f= - f_\leftarrow$
(see also fig.~\ref{flow.fig}).  For higher rotation speed one expects
that the material moves homogeneously and there are only small
fluctuations of $\Delta f$. The impulses in the upper part of
fig.~\ref{phiovert.fig} show the cumulative positive material flow
between each two consecutive times $t_i^{(\Delta f = 0)}$ and
$t_{i+1}^{(\Delta f = 0)}$ when $\Delta f=0$ divided by the negative
flow $f_\leftarrow$, i.e.
\begin{equation}
f_{cum}\left(\frac{t_{i+1}-t_{i}}{2} \right) = \frac{1}{f_\leftarrow} 
   \int_{t_i^{(\Delta f = 0)}}^{t_{i+1}^{(\Delta f = 0)}} \Delta f dt~.
  \label{cummflow}
\end{equation}
These values give a quantitative measure for the irregularity of the
flow. One can see that for low $\Omega$ one observes fewer but much
bigger distinguishable avalanches than for high angular velocity.
For low angular velocity $\Omega$ one finds a strong
correlation between avalanches (top figure~\ref{phiovert.fig}) and sudden
decreases of the inclination $\Theta$ (bottom figure).

\begin{figure}[ht]
\centerline{\psfig{figure=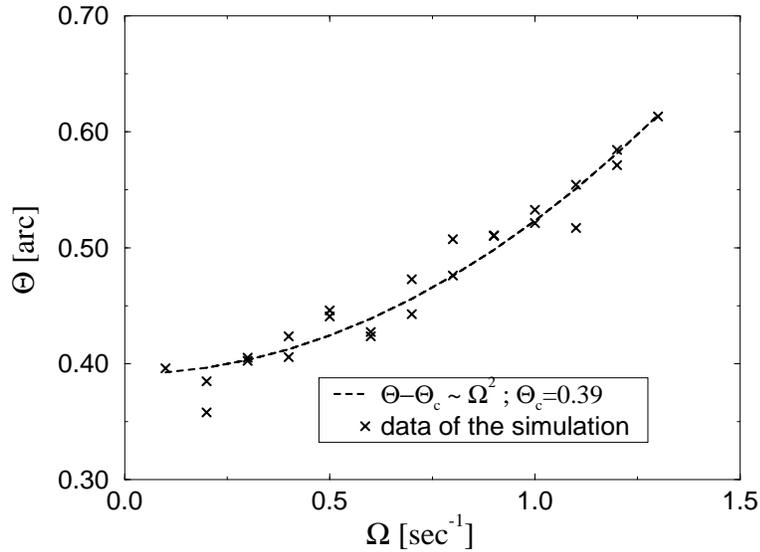,width=10cm,angle=270}}
\vspace{0.5cm}
\caption{\em The inclination $\Theta$ of the material surface over 
  the angular velocity $\Omega$. The dotted line displays the function
  which has been measured experimentally by
  \nobreak{Rajchenbach}~[6].}
%\cite{Rajchenbach:1990}.}
\label{inclination.fig}
\end{figure}

The inclination of the surface of the material is a function
of the angular velocity. Rajchenbach~\cite{Rajchenbach:1990}
determined the law experimentally to be
\begin{equation}
\Theta-\Theta_c \sim \Omega^2~,
  \label{inclination.eq}
\end{equation}
where $\Omega_c$ is a constant. For each angular velocity $\Omega$ we
measured in the simulation the inclination of the surface $\Theta$.
Fig.~\ref{inclination.fig} shows the inclination $\Theta$ of the
material surface over the angular velocity $\Omega$ and the function
in eq.~(\ref{inclination.eq}). The numerical data points are located
at both sides of the experimental curve. Although there are not enough
data points to confirm the experimental results, we can show at least
that the experimental and theoretical results do not contradict each
other. The measured value for the critical angle is $\Theta_c =
22.3^o$. Typical values for the experimentally measured angle of
repose lie between $\Theta_c = 20^o$ and $\Theta_c = 30^o$,
Bretz~et~al. for instance found $\Theta_c \approx 25^o$
\cite{Bretz:1992}. Values for the critical angle found in earlier
simulation \cite{PoeschelBuchholtz:1993CSF} are $\Theta_c \approx 8^o$
for circular particles and $\Theta_c \approx 18^o$ for non--circular
compound particles~\cite{PoeschelBuchholtz:1993CSF}.

\section{Conclusion}
We presented a model to simulate the translational and rotational
movement of complex particles in a rotating drum. The particles
consist of triangles connected to each other by beams. Energy
dissipation is caused by inner degrees of freedom. Our model is able
to predict the angle of the surface of the granular material with the
horizontal line with higher accuracy than other models. The angle
given by our model is in the same range as experimental values and
higher than the angle predicted by models using circular disks or particles
consisting of circles. We think that this advantage of the model is
caused by the use of particles with the more realistic polygonal shape
which allows to take steric effects into account. One drawback of our
model is its slowness which is due to the complexity of the
numerical calculation. As also observed in experiments and in other
simulations we found two flow regimes, a stick--slip motion of the
grains for low drum frequencies and a continuous flow regime for high
frequencies. Our data fit well with the power law between the surface
flow and the surface angle found experimentally by Rajchenbach.
Unfortunately, they are not yet sufficient to calculate the precise
exponent. Further optimization  and more simulation runs have to be
performed.

\ack We thank H.~J.~Herrmann, S.~Melin, S.~Schwarzer and F.~Spahn for
helpful discussions, and S.~Seefeld for the animated visualization of
the data.

\vfill
\end{document}